\documentclass[conference]{IEEEtran}
\IEEEoverridecommandlockouts
\usepackage{algorithm}
\usepackage{cite}
\usepackage{dblfloatfix}
\usepackage{float}
\usepackage{amsmath,amssymb,amsfonts}
\usepackage{mathtools}
\usepackage{algorithm}
\usepackage[noend]{algpseudocode}
\usepackage{graphicx}
\graphicspath{{./Figures/}}
\usepackage{textcomp}
\usepackage{siunitx}
\usepackage{url}
\usepackage{array,multirow}
\usepackage[table]{xcolor}
\usepackage{xcolor,colortbl}
\def\BibTeX{{\rm B\kern-.05em{\sc i\kern-.025em b}\kern-.08em
T\kern-.1667em\lower.7ex\hbox{E}\kern-.125emX}}

\begin{document}	
\title{Maximum Power Reference Tracking Algorithm for Power Curtailment of Photovoltaic Systems}

\author{\IEEEauthorblockN{Victor Paduani, Lidong Song, Bei Xu, Dr. Ning Lu}
\IEEEauthorblockA{Department of Electrical and Computer Engineering} 
North Carolina State University, Raleigh, NC\\

\thanks{This research is supported by the U.S. Department of Energy's Office of Energy Efficiency and Renewable Energy (EERE) under the Solar Energy Technologies Office Award Number DE-EE0008770.}
}
\maketitle

\begin{abstract}
\textbf{This paper presents an algorithm for power curtailment of photovoltaic (PV) systems under fast solar irradiance intermittency. Based on the Perturb and Observe (P\&O) technique, the method contains an adaptive gain that is compensated in real-time to account for moments of lower power availability. In addition, an accumulator is added to the calculation of the step size to reduce the overshoot caused by large irradiance swings. A testbed of a three-phase single-stage, 500 kVA PV system is developed on the  OPAL-RT eMEGAsim real-time simulator. Field irradiance data
and a regulation signal from PJM (RTO) are used to compare the performance of the proposed method with other techniques found in the literature. Results indicate an operation with smaller overshoot, less dc-link voltage oscillations, and improved power reference tracking capability.}
\end{abstract}


\section{Introduction}
In grids with high penetration of photovoltaic (PV) systems, there is an increasing need for employing inverters with superb controllability to provide grid support functions (GSFs). IEEE Standard 1547-2018 \cite{ieee2018ieee} imposes new operational requirements on smart inverters to accelerate technological developments for enabling PV systems at all levels to provide high quality grid services. Main GSFs of interest include power curtailment, voltage and frequency droop control, fast frequency response, and reactive power regulation \cite{gevorgian2016advanced}. 

Although maximum power point tracking (MPPT) algorithms for improving irradiance tracking are well researched in the literature, power curtailment algorithms remained under-examined until the last decade. 
In \cite{wandhare2014precise}, Wandhare and Agarwal introduce a power curtailment algorithm for single-staged PV systems, in which the dc-link voltage is controlled by the inverter in fixed voltage steps based on the perturb and observe (P\&O) algorithm. Meanwhile, other power limiting strategies based on proportional-integral controllers 
are introduced by Cao \textit{et al.} in \cite{cao2013two}.

In \cite{sangwongwanich2017benchmarking}, Sangwongwanich \textit{et al.} compare the P\&O- algorithm with the PI-based approach for constant power generation (CPG) applications and they show that the CPG--P\&O method achieved the highest robustness at the cost of worse dynamic response. 
In \cite{tafti2018adaptive}, Tafti \textit{et al.} introduced an adaptive algorithm in which the step size during transients is proportional to the difference between the output power and the power reference. Results confirmed less overshoot and faster convergence during irradiance changes. However, as the error increases drastically during sudden irradiance drops caused by passing clouds, the calculated variable voltage step size from the adaptive algorithm may become too big and cause large oscillations in the dc-link voltage, adding unwanted ripple to the output power and extra stress to circuit components.

Furthermore, despite the successful classification between transient and steady-state conditions, the issue of large power overshoots remains as noted in  \cite{narang2019algorithm}. In normal operation, when solar irradiance increases, the PV output power also increases. Thus, as the system approaches its setpoint, the step size becomes smaller and smaller because of the diminishing power tracking error. This leads to a slower system response until the error becomes large again. Hence, overshoots are inevitable. Consequently, there is a trade-off between the size of the adaptive gain and the power overshoot.

\begin{figure*}[!t]
	\centering
	\includegraphics[width=0.75\textwidth]{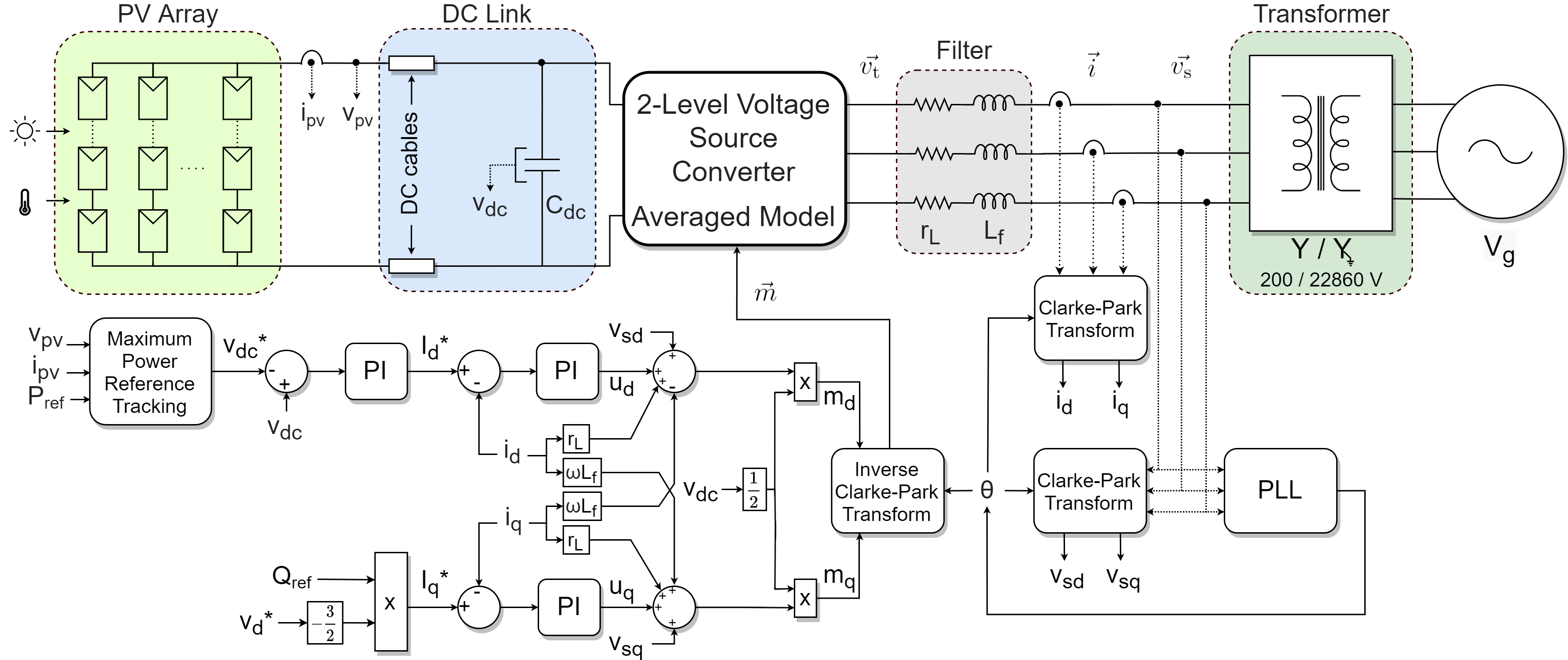}
	\caption{Circuit and control system block diagrams of a grid-scale PV system.}
	\label{maindiagram}
\end{figure*}

To address the aforementioned issues, in this paper we propose two independent solutions that do not require external irradiance or temperature sensors. First, we introduce a compensation factor that is used to adjust the adaptive gain used to calculate the voltage step size in real-time. Second, we use an accumulator, which serves as a memory for tracking the most recent changes in PV output power, so the algorithm can maintain a large voltage step size during fast irradiance changes when the PV is operating closer to its power reference.

The rest of the paper is organized as follows: Section II introduces the grid-scale PV system model and the proposed algorithms; Section III describes the simulation environment; Section IV presents the simulation results; Section V concludes the paper.



\section{Methodology}
In this section, we will present models and control methods used in grid-scale PV systems and introduce the improvements proposed to enhance the power curtailment algorithm.
\subsection{PV System Modeling}
Figure 1 shows the circuit diagram of a single-stage, centralized inverter found in a grid-scale PV farm.
The PV array is modeled with a 5-parameter model from \cite{PVarray}, whereas 
the inverter is modeled with an averaged model of a two-level voltage source converter (VSC) developed in \cite{yazdani2010voltage}. 

The terminal voltage of the two-level VSC before its output filter, $\vec{v_{\mathrm{t}}}$, can be calculated as
\begin{equation}
\vec{v_{\mathrm{t}}} = \frac{v_{\rm{dc}}}{2}\vec{m}
\end{equation}
where $\vec{m}$ is the modulation signal and $v_{\mathrm{dc}}$ is the dc bus voltage. 

Then, using KCL, the secondary voltage of the transformer, ($\vec{v_{\mathrm{s}}}$), can be expressed as
\begin{equation}
\vec{v_{\mathrm{s}}} = \vec{v_{\mathrm{t}}} - r_{\mathrm{L}}\vec{i} - L_{\mathrm{f}}\frac{d\vec{i}}{dt}
\end{equation}
where $i$ is the output current of the converter; $r_\mathrm{L}$ and $L_\mathrm{f}$ is the filter resistance and inductance, respectively.

As shown in the control system block diagram in Fig. 1, the PV farm is operated in the grid-following mode, and the PV inverter controller will regulate the output current so that the power supplied from the dc-link to the grid follows the real and reactive power reference signals, $P_\mathrm{{ref}}$ and $Q_\mathrm{{ref}}$. Clarke-Park transformation is used to convert the 3-phase voltage measurements to dq0 coordinates whereas a phase-locked loop (PLL) is used to extract the phase of the grid.

In dq0 coordinates, a phasor $\vec{x}$ can be defined as $x_{\mathrm{d}} + jx_{\mathrm{q}}$, and its derivative can be represented as
\begin{equation}
\frac{d\vec{x}}{dt} = \frac{dx_{\mathrm{d}}}{dt} + j\frac{dx_{\mathrm{q}}}{dt} + j\omega(x_{\mathrm{d}}+jx_{\mathrm{q}})
\end{equation}
where $\omega$ is the nominal angular frequency of the system.

By separating the real and imaginary parts, (3) can be rewritten as
\begin{equation}
L\frac{di_{\mathrm{d}}}{dt} = m_{\mathrm{d}}\frac{v_{\mathrm{dc}}}{2}+ \omega L_{\mathrm{f}}i_{\mathrm{q}} - r_{\mathrm{L}}i_{\mathrm{d}} - v_{\mathrm{sd}}
\end{equation}
\begin{equation}
L\frac{di_{\mathrm{q}}}{dt} = m_{\mathrm{q}}\frac{v_{\mathrm{dc}}}{2}- \omega L_{\mathrm{f}}i_{\mathrm{d}} - r_{\mathrm{L}}i_{\mathrm{q}} - v_{\mathrm{sq}}
\end{equation}
where $i_d$ and $i_q$, $v_{\mathrm{sd}}$ and $v_{\mathrm{sq}}$, $m_\mathrm{d}$ and $m_\mathrm{q}$, are the direct and quadrature components of $i$, $v_s$, and $m$, respectively. 
To decouple the currents and remove $v_{\rm{sd}}$ and $v_{\rm{sq}}$ from the control system, $m_d$ and $m_q$, can be calculated as
\begin{equation}
m_{\mathrm{d}} = \frac{2}{v_{\mathrm{dc}}}\Big(v_{\mathrm{sd}} - \omega L_{\mathrm{f}}i_{\mathrm{q}} + r_{\mathrm{L}}i_{\mathrm{d}} + u_{\mathrm{d}}\Big)
\end{equation}
\begin{equation}
m_{\mathrm{q}} = \frac{2}{v_{\mathrm{dc}}}\Big(v_{\mathrm{sq}} +\omega L_{\mathrm{f}}i_{\mathrm{d}} + r_{\mathrm{L}}i_{\mathrm{q}} + u_{\mathrm{q}}\Big)
\end{equation}
where $u_{\mathrm{d}}$ and $u_{\mathrm{q}}$ are the outputs of the current controllers. Notice that the filter resistance is also included as a feed-forward term to improve the control performance. The three-phase modulation signal, $\vec{m}$, is then generated by $m_d$ and $m_q$ using the inverse Clarke-Park transformation. 


\subsection{Power Curtailment Algorithm}



\begin{figure}[htb]
	\centerline{\includegraphics[width=0.4\textwidth]{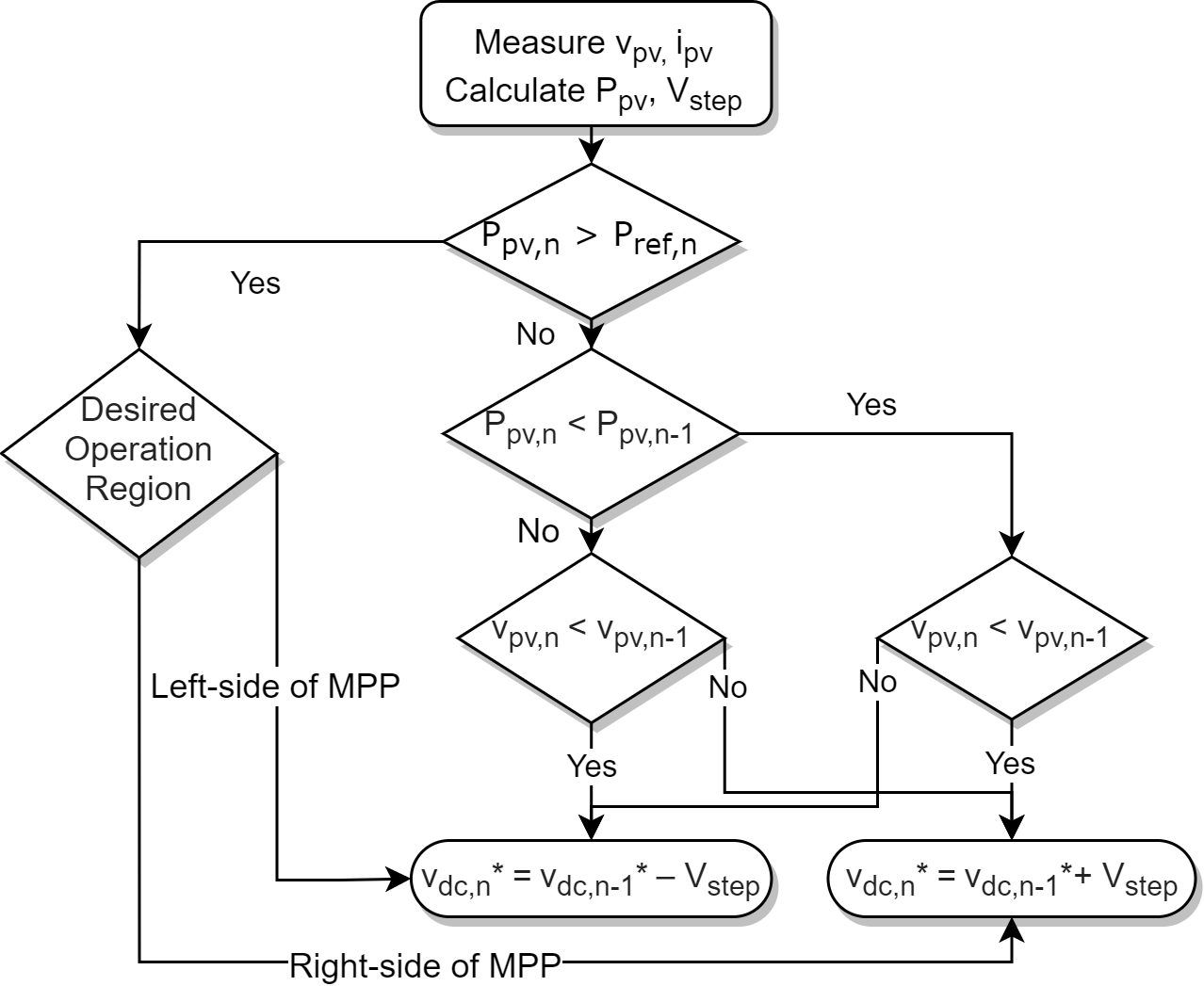}}
	\caption{Flowchart of a basic P\&O power curtailment algorithm.}
	\label{flow1}
\end{figure}

The principle of operation of the P\&O technique is well established in the literature. Based on its strategy, a power curtailment algorithm can be created. According to the flowchart in Fig. \ref{flow1}, when the output power of the panels, $P_\mathrm{{pv}}$, is below a given reference, $P_\mathrm{{ref}}$, the system operates as the P\&O method; when $P_\mathrm{{pv}}$ is above $P_\mathrm{{ref}}$, the voltage reference, $v^{*}_\mathrm{dc}$, is adjusted to curtail $P_\mathrm{{pv}}$. Whether the current operation point is on the right side or the left side of the maximum power point (MPP) can be determined by increasing or decreasing $v^{*}_\mathrm{dc}$, respectively, with a step size of $V_\mathrm{step}$, whenever $P_\mathrm{{pv}}$ is above $P_\mathrm{{ref}}$. In a single-stage PV system, because of inherent limitations of inverters \cite{paduani2019design}, the dc-link voltage, $V_\mathrm{dc}$, should always be kept above a minimum voltage threshold, so operation on the right side of the MPP is necessary. 



The algorithm's convergence speed is primarily determined by $V_\mathrm{step}$ at each iteration. Ideally, $V_\mathrm{step}$ should be small in steady-state operation to reduce power oscillations around $P_\mathrm{ref}$, and large during transients for a rapid response to changes in $P_\mathrm{ref}$ or irradiance. As introduced in \cite{tafti2018adaptive}, a fast convergence during transient operation can be achieved by making $V_\mathrm{step}$ a function of the error from the power reference $P_\mathrm{{err}}$, so we have 
\begin{align}
\begin{split}
	V_{\mathrm{step,}n} &= \alpha \times V_{\mathrm{step}}^{\mathrm{min}}+(1-\alpha)\times K_{\mathrm{tr}} |P_{\mathrm{err,}n}|\\
	P_{\mathrm{err,}n} &= P_{n} - P_{\mathrm{ref,}n}
\end{split}
\end{align}
where $\alpha$ is used to switch between the power error consideration in the transient operation and the fixed minimum step change in the steady-state operation, and $K_\mathrm{tr}$ is a constant gain for considering $P_\mathrm{{err}}$.

\subsection{Adaptive Gain Adjustment}
There is a major limitation when using (8) to calculate $V_\mathrm{step}$. When irradiance suddenly drops to a level where it is impossible for the PV system to converge to $P_\mathrm{ref}$,  $V_\mathrm{step}$ will be too large, causing the DC link voltage to change rapidly. To attenuate this issue, in \cite{tafti2018adaptive}, a mode classification method for separating steady-state and transient modes is used to switch $V_\mathrm{step}$ to a smaller value when irradiance is low. However, the tuning method for the operation mode classification is not clearly defined, and if it is not well designed, the issue can be further aggravated. Besides, irradiance variations happen very often amid the region of lower irradiance, which causes the system to switch to transient mode frequently.

Therefore, in this paper, we introduce a compensation factor for adjusting $K_\mathrm{{tr}}$ in real-time to provide a robust solution to this issue. First, we calculate the  moving average of the power output of the system at time interval $n$ over a window of length $N$, as
\begin{equation}
\overline{P_{\mathrm{pv}\mathrlap{,n}}}_{\hphantom{,\mathrm{n}}} = \frac{1}{N}\sum_{i=0}^{N-1}P_{\mathrm{pv,}n\text{-}i}
\end{equation}


Then, by comparing $\overline{P_{\mathrm{pv}\mathrlap{,n}}}_{\hphantom{,\mathrm{n}}}$ with $P_\mathrm{ref}$,  $K_{tr}$ is calculated as
\begin{equation}
K_{\mathrm{tr}} = Max\Bigg[C_{\mathrm{min}}K_{\mathrm{base}}, K_{\mathrm{base}} \bigg(\dfrac{\overline{P_{\mathrm{pv}\mathrlap{,n}}}_{\hphantom{,\mathrm{n}}}}{P_{\mathrm{ref}}}\bigg)^{2}\Bigg]
\label{Kupdate}
\end{equation}
where $C_{\mathrm{min}}$ defines the lower boundary related to the minimum $K_{\mathrm{tr}}$ acceptable, and $K_{\mathrm{base}}$ is the original gain.

\begin{figure}[hbt]
	\centerline{\includegraphics[width=0.5\textwidth]{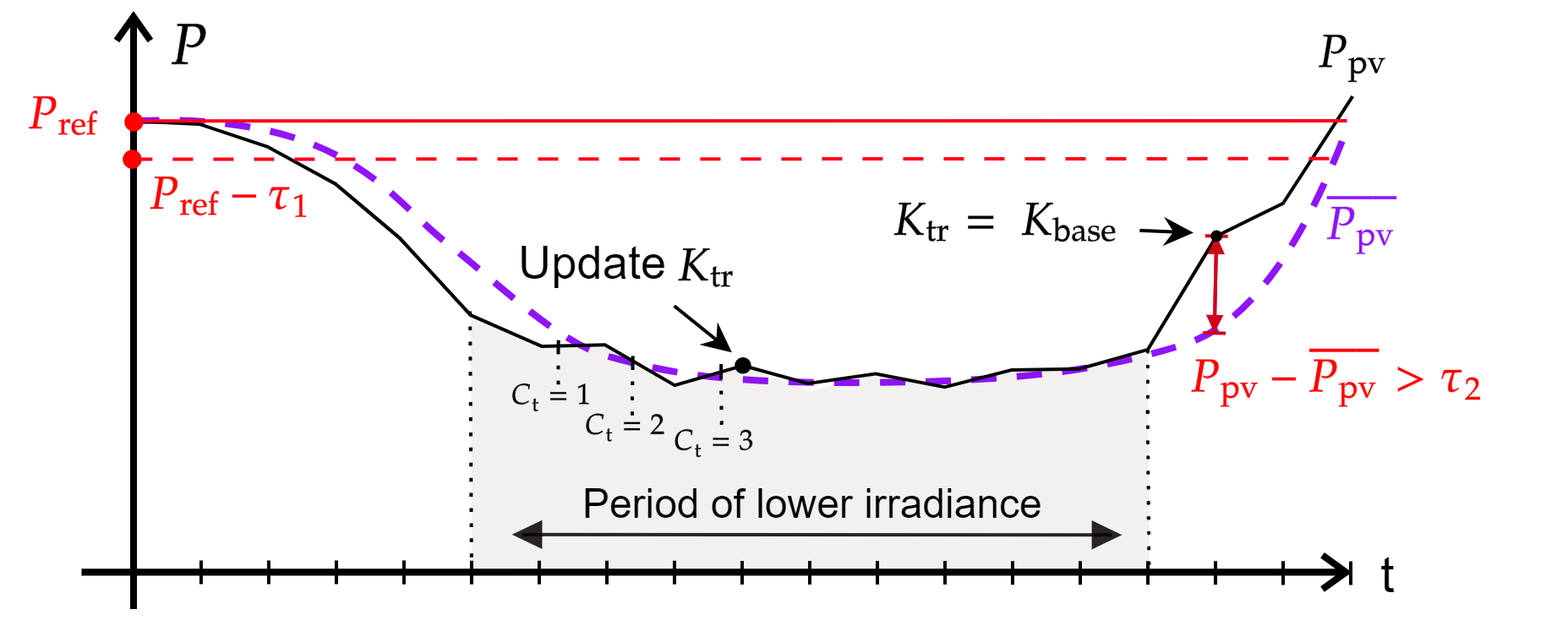}}
	\caption{Example of the adaptive $K_{tr}$ adjustment.}
	\label{kadjust}
\end{figure}

The pseudo code for adaptively adjusting $K_\mathrm{{tr}}$ is shown in Algorithm 1. When $P_{\mathrm{pv}}<P_\mathrm{ref}$, the steady-state operation point is the MPP corresponding to the solar irradiance, which is varying constantly. As illustrated in Fig. \ref{kadjust}, to determine whether or not an MPP is reached, a counter, $C_{\mathrm{t}}$, is used to count the number of continuous crossings between $P_{\mathrm{pv}}$ and $\overline{P_{\mathrm{pv}\mathrlap{,n}}}_{\hphantom{,\mathrm{n}}}$.  If $P_{\mathrm{pv,n}}$ crosses $\overline{P_{\mathrm{pv}\mathrlap{,n}}}_{\hphantom{,\mathrm{n}}}$,  $C_{\mathrm{t}}$ is incremented; otherwise $C_{\mathrm{t}}$ is reset to zero. Once the counter threshold, $C_{\mathrm{t,max}}$, is reached, $K_{\mathrm{tr}}$ will be adjusted. In this paper, we set $C_{t,\mathrm{max}} = 3$ to balance the adjustment speed of $K_\mathrm{{tr}}$ and the MPP detection accuracy. Once adjusted, $K_{\mathrm{tr}}$ will maintain its new value. If $C_{\mathrm{t}}=C_{\mathrm{t,max}}$ again, another adjustment will be made.  If irradiance increases quickly causing a large increase in $P_{\mathrm{pv,}n}$ compared with $P_\mathrm{ref}-\tau_{1}$ or entering the control deadband so that  $|P_{\mathrm{pv,}n} - \overline{P_{\mathrm{pv}\mathrlap{,n}}}_{\hphantom{,\mathrm{n}}}| > \tau_{2}$, $K_{\mathrm{tr}}$ will be reset to $K_{\mathrm{base}}$, as shown in operation 6 of Algorithm 1 and Fig. \ref{kadjust}.
\begin{algorithm}
\caption{Adaptive $K_\mathrm{{tr}}$ adjustment}
\begin{algorithmic}[1]
    \small
    \State Calculate $\overline{P_{\mathrm{pv}\mathrlap{,n}}}_{\hphantom{,\mathrm{n}}}$ \Comment{(9)}
    \If{$\big(P_{\mathrm{pv,}n}-\overline{P_{\mathrm{pv}\mathrlap{,n}}}_{\hphantom{,\mathrm{n}}}\big)\big(P_{\mathrm{pv,}n\text{-}1}-\overline{P_{\mathrm{pv}\mathrlap{,n}}}_{\hphantom{,\mathrm{n}}}\big) > 0$}
    \State{$C_{\mathrm{t}}=0$}
    \Else
    \State{$C_{\mathrm{t}}\mathrel{+}=1$}
    \EndIf
    \If{$P_{\mathrm{pv,}n} > P_\mathrm{ref}-\tau_{1}$ \textbf{or} $|P_{\mathrm{pv,}n} - \overline{P_{\mathrm{pv}\mathrlap{,n}}}_{\hphantom{,\mathrm{n}}}| > \tau_{2}$}
    \State{$K_{\mathrm{tr,}n}= K_{\mathrm{base}}$}
    \ElsIf{$C_{\mathrm{t}} \geq C_{\mathrm{t,max}}$}
        \State{Update $K_{\mathrm{tr}}$}\Comment{(10)}
    \Else
        \State{$K_{\mathrm{tr,}n} = K_{\mathrm{tr,}n\text{-}1}$} 
    \EndIf
\end{algorithmic}
\end{algorithm}

\subsection{Accumulator for Overshoot Suppression}
Overshoot is still a critical issue found in P\&O-based power curtailment algorithms (see Fig. 16c in \cite{sangwongwanich2017benchmarking}). The common cause of an overshoot is the rapid increase of solar irradiance after a passing cloud. If a controller cannot curtail $P_{\mathrm{pv}}$ fast enough, an overshoot over  $P_{\mathrm{ref}}$ will occur. Recall when we adjust $V_{\mathrm{step}}$ based on $P_\mathrm{{err}}$ in (8), $V_{\mathrm{step}}$ becomes smaller as $P_{\mathrm{pv},n}$ moves closer to $P_\mathrm{ref}$. Thus, with a reducing $V_{\mathrm{step}}$ and an increasing solar irradiance, $P_{\mathrm{pv}}$ will overshoot $P_{\mathrm{ref}}$ and the overshoot can be long-lasting unless $V_{\mathrm{step}}$ can be increased accordingly. To suppress the overshoots, we introduce an accumulator, $\gamma$, into the $V_{\mathrm{step}}$ calculation. At the beginning of each iteration, the moving average of the absolute error variation, $\overline{\Delta P_{\mathrm{err}}}_{,n}$, is calculated by 
\begin{align}
\begin{split}
		\Delta P_{\mathrm{err,} n} &= |P_{\mathrm{err,} n}| - |P_{\mathrm{err,} n-1}|\\
		\overline{\Delta P_{\mathrm{err}\mathrlap{,n}}}_{\hphantom{,\mathrm{n}}} & = \frac{1}{M}\sum_{i=0}^{M-1}\Delta P_{\mathrm{err,} n-i}
\end{split}
\end{align}

Then, if $\overline{\Delta P_{\mathrm{err}\mathrlap{,n}}}_{\hphantom{,\mathrm{n}}}>0$ and $P_{\mathrm{pv},n} > P_\mathrm{ref}$, marking the detection of an overshoot, we will add $\gamma_n$ to the $V_{\mathrm{step,n}}$ calculated from (8). If there is no overshoot detected, but variations in $P_{\mathrm{pv},n}$ have been positive in the previous two iterations, then $\gamma_n$ is increased by (\ref{ddP2}). Otherwise, $\gamma_n$ is multiplied by a resetting rate, $\lambda_{\mathrm{r}}$. 
\begin{equation}
\gamma_{n} = \gamma_{n-1} + K_{\mathrm{acc}}\times K_{\mathrm{base}} \times |P_{\mathrm{err,}n}|
\label{ddP2}
\end{equation}
The overshoot suppression is presented in Algorithm 2. 
\begin{algorithm}
\caption{Accumulator $\gamma_{n}$}
\begin{algorithmic}[2]
    \small
    \State Calculate $\overline{\Delta P_{\mathrm{err}\mathrlap{,n}}}_{\hphantom{,\mathrm{n}}}$ \Comment{(11)}
    \If{$\overline{\Delta P_{\mathrm{err}\mathrlap{,n}}}_{\hphantom{,\mathrm{n}}} > 0$ \textbf{and} $P_{\mathrm{pv},n} > P_\mathrm{{ref}}$}
    \State{$V_{\mathrm{step}} = V_{\mathrm{step}} + \gamma_{n}$}
    \ElsIf{$P_{\mathrm{pv},n} > P_{\mathrm{pv},n\text{-}1} > P_{\mathrm{pv},n\text{-}2}$}
    \State{Increase $\gamma_{n}$}\Comment{(12)}
    \Else
    \State{$\gamma_{n} = \lambda_{\mathrm{r}}\gamma_{n}$}
    \EndIf
\end{algorithmic}
\end{algorithm}

\section{Simulation Results}
To test the proposed adaptive gain adjustment and accumulator algorithms, we set up a testbed of a 500 kVA PV system on an OPAL-RT real-time simulation platform. One-second irradiance data collected from a 1.04 MW solar farm by EPRI \cite{EPRIdata} is used. To demonstrate the power following capability in a realistic application, we apply a unidirectional 200 kW PJM regulation signal to the estimated available power of the system to generate the power reference command. The estimated power is obtained by applying a low-pass filter to the irradiance. The simulation timestep is \SI{100}{\micro\s}. To account for measurement noise due to sensor imperfections, a Gaussian noise was added to $v_{\mathrm{pv}}$, $i_{\mathrm{pv}}$, and $v_{\mathrm{dc}}$, creating a signal-to-noise-ratio of 71 dB. The rest of the simulation parameters are listed in Table \ref{parameters}.

\renewcommand{\arraystretch}{1.05}
\begin{table}[ht]
	\caption{Simulation Parameters}
	\begin{center}
		\begin{tabular}{|>{\columncolor[gray]{0.85}} c|c|c|}
			\hline
			
			&Power&\SI{612}{\kilo\watt}\\\cline{2-3}
			&Module&CS6P-250P\\\cline{2-3}
			&Size (parallel$\times$series) &153 $\times$ 16\\\cline{2-3}
			\multirow{-4}{*}{PV Array}&V$_{\mathrm{mpp}}$, I$_{\mathrm{mpp}}$& \SI{481.6}{\volt}, \SI{1270}{A}\\
			\hline\hline
			&Power, Frequency&\SI{500}{\kilo VA}, \SI{60}{\hertz} \\\cline{2-3}
			&L$_{\mathrm{f}}$, r$_{\mathrm{L}}$&\SI{100}{\micro\henry}, \SI{3}{\milli\ohm}\\\cline{2-3}
			&C$_{\mathrm{dc}}$& \SI{5000}{\micro\farad}\\\cline{2-3}
			&PI (v$_{\mathrm{dc}}$)& K$_{\mathrm{p}}$ = 1, K$_\mathrm{{i}}$ = 250\\\cline{2-3}
			\multirow{-5}{*}{Inverter}&PI (i$_{\mathrm{d}}$, i$_{\mathrm{q}}$)& K$_{\mathrm{p}}$ = 0.7, K$_{\mathrm{i}}$ = 50\\
			\hline\hline
			&Power, Frequency&\SI{500}{\kilo VA}, \SI{60}{\hertz} \\\cline{2-3}
			&V$_{\mathrm{LL}}$ (rms)&\SI{200}{\volt} / \SI{22.86}{\kilo\volt}\\\cline{2-3}
			&X$_{\mathrm{leak}}$, r$_{\mathrm{loss}}$ (pu)& 0.06, 0.0024\\\cline{2-3}
			&L$_{\mathrm{m}}$, r$_{\mathrm{m}}$ (pu)& 200, 200\\\cline{2-3}
			\multirow{-5}{*}{Transformer}&Core type& Three-limb\\
			\hline\hline
			&Sampling frequency& \SI{5}{\hertz} \\\cline{2-3}
			&K$_{\mathrm{base}}$, K$_{\mathrm{acc}}$, C$_{min}$ & 0.00006, 0.3, 0.2 \\\cline{2-3}
			&$V_{\mathrm{step}}^{\mathrm{min}}$ / $V_{\mathrm{step}}^{\mathrm{max}}$ & 0.3 / 12 V\\\cline{2-3}
			&$\tau_{2}, \tau_{1}$ & 7.5 / \SI{10}{\kilo\watt}\\\cline{2-3}
			\multirow{-4}{*}{MPRT}&N, M, $C_{\mathrm{t,max}}$, $\lambda_{r}$& 4, 3, 3, 0.5\\
			\hline
		\end{tabular}
	\end{center}
	\label{parameters}
\end{table}
The performance of three control methods are compared. In method 1, $V_{\mathrm{step}}=\SI{0.3}{\volt}$ in steady-state and $V_{\mathrm{step}}=\SI{4}{\volt}$ in transient operation. In method 2, $V_{\mathrm{step}}=\SI{0.3}{\volt}$ in steady-state, and (8) proposed by \cite{tafti2018adaptive} is used to compute $V_{\mathrm{step}}$ during transients. Method 3 is the proposed algorithm. 

\subsection{Overshoot Suppression}
\begin{figure}[htb]
	\centerline{\includegraphics[width=0.45\textwidth]{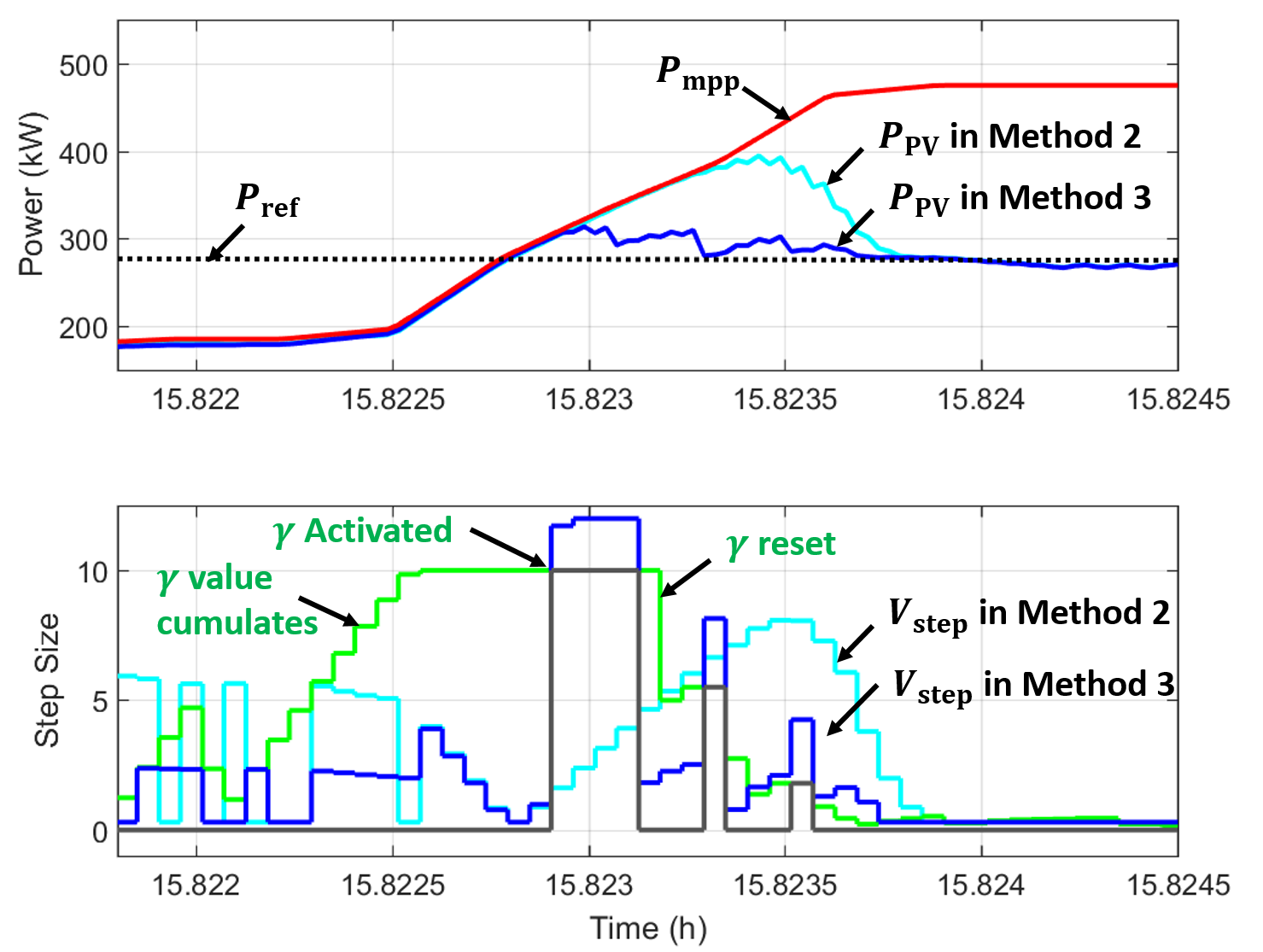}}
		\caption{Overshoot response comparison.}
	\label{overshoot}
\end{figure}

Figure \ref{overshoot} demonstrates the efficacy of employing an accumulator mechanism for calculating $V_{\mathrm{step}}$ during large irradiance variations. Because the initial irradiance increase makes $P_{\mathrm{pv}}$ approach $P_{\mathrm{ref}}$, $V_{\mathrm{step}}$ calculated by (8) (method 2) will continuously decrease even when the irradiance is still rapidly increasing. This can cause a large overshoot. In method 3, $\gamma$ value accumulates when irradiance increases. Once an overshoot is detected, the accumulator is activated and its value is added to $V_{\mathrm{step}}$, providing a faster response to suppress the overshoot, as shown in the first plot in Fig. \ref{overshoot}. Because the resetting rate of the accumulator is greater than zero, it can be activated multiple times throughout one overshoot event, as shown in the second plot of Fig. \ref{overshoot}. 

\subsection{Smoothing DC-link Voltage Oscillations}
\begin{figure}[htb]
	\centerline{\includegraphics[width=0.4\textwidth,height=50mm]{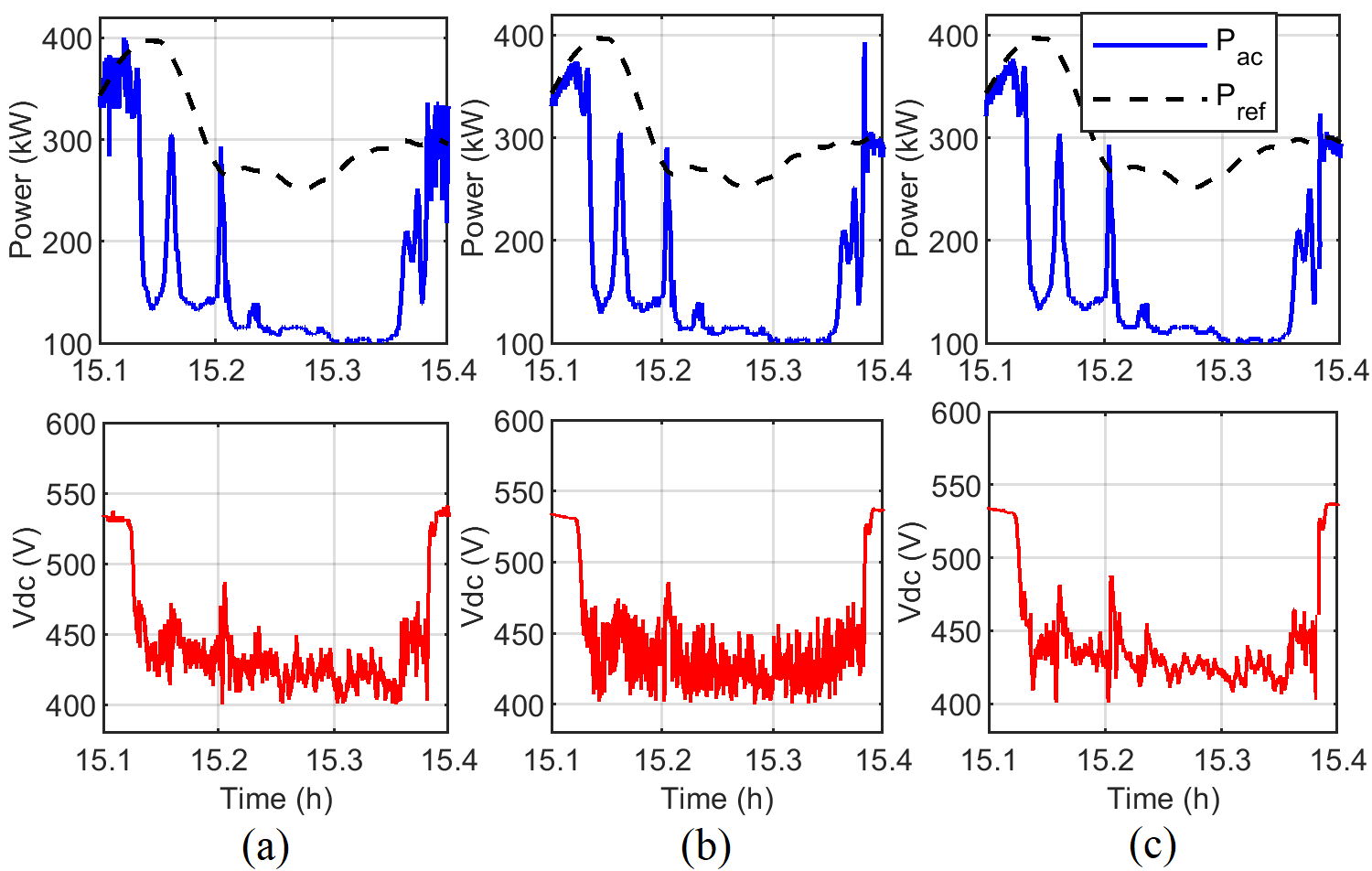}}
	\caption{Comparison of DC-link voltage oscillations: (a) method 1, (b) method 2, (c) method 3.}
	\label{dclink}
\end{figure}

Figure \ref{dclink} displays the operation of each method during periods of low irradiance. By adjusting $K_{\mathrm{tr}}$ in real-time, method 3 can reduce the oscillations in $v_{\mathrm{dc}}$ during low-irradiance conditions when $P_{\mathrm{ref}}>P_{\mathrm{mpp}}$, making the system more stable and causing less wear-and-tear. In Table \ref{table2}, we use  the cumulative voltage oscillations ($\sum |v_{\mathrm{dc}(t)}-v_{\mathrm{dc}(t)}|$) in three consecutive days to compare the performance of each method at different irradiance levels. Note that lower $v_{\mathrm{dc}}$ corresponds to lower irradiance cases. It can be seen clearly that method 3 outperforms methods 1 and 2 in low irradiance cases.

\renewcommand{\arraystretch}{1}
\begin{table}[ht]
	\caption{DC-link Voltage Oscillations}
	\begin{center}
		\begin{tabular}{|>{\columncolor[gray]{0.85}}c|c|c|c|}
		\hline
		    Method &  $V_{\mathrm{dc}} < \SI{450}{\volt}$ & $450 \leq V_{\mathrm{dc}} < \SI{500}{\volt}$ & $V_{\mathrm{dc}} \geq \SI{500}{\volt}$\\		\hline
			 1 & 121k & 69k&417k\\
			\hline
			 2 & 149k & 63k&171k\\
			\hline
			 3 & 77k & 50k&171k\\
			\hline
		\end{tabular}
	\end{center}
	\label{table2}
\end{table}
			
\subsection{Power Reference Tracking}
\begin{figure}[htb]
	\centerline{\includegraphics[width=0.4\textwidth]{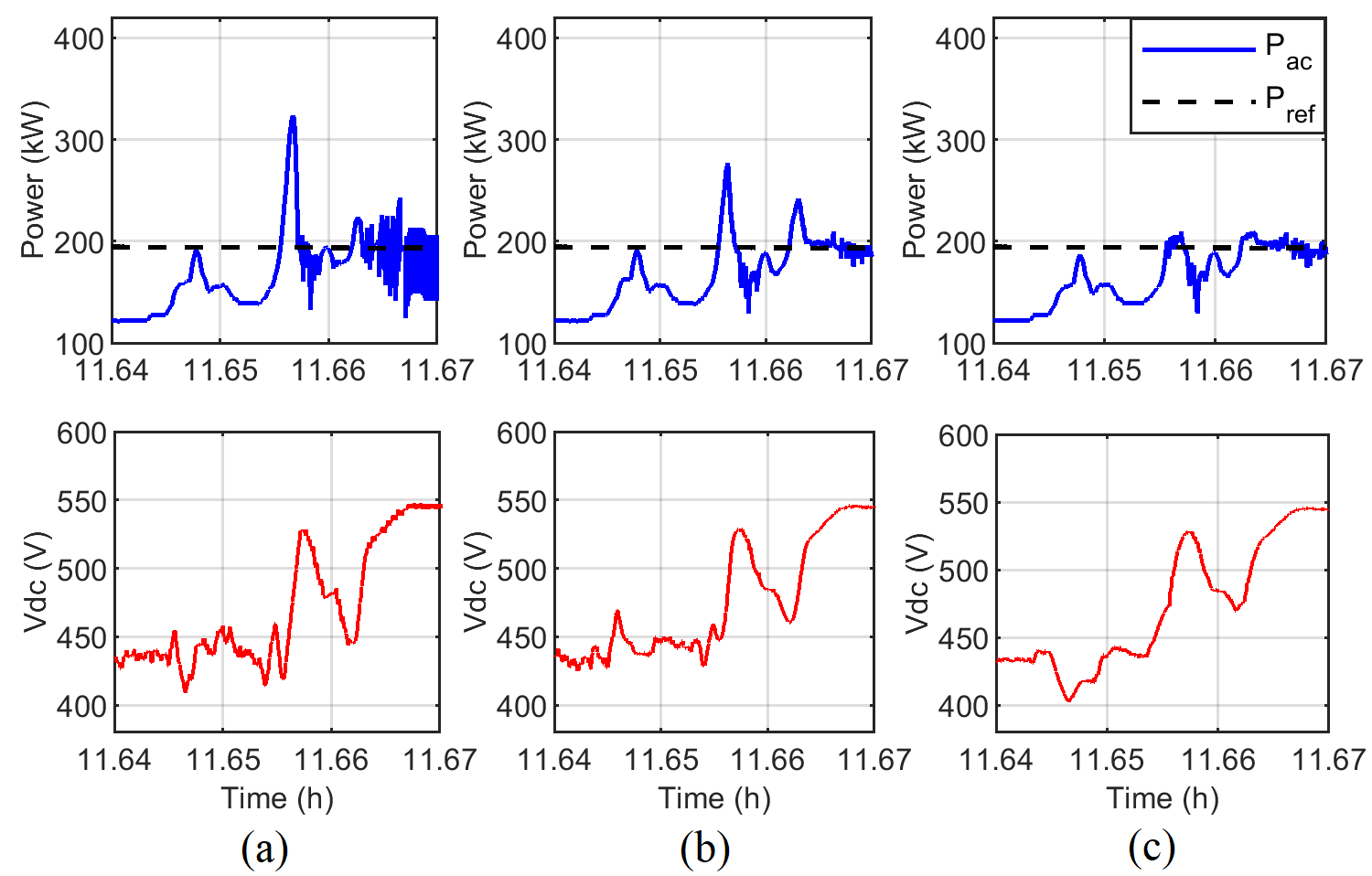}}
	\caption{$P_\mathrm{{ref}}$ tracking comparison: (a) method 1, (b) method 2, (c) method 3.}
	\label{Pout}
\end{figure}

The performance of the three methods when tracking $P_\mathrm{{ref}}$ for providing regulation signals is shown in Fig. \ref{Pout}.  Method 1 presents large oscillations in transient responses because of its fixed transient $V_\mathrm{{step}}$. Method 2 can mitigate the oscillations, but it is susceptible to overshoots caused by large irradiance changes. Method 3, equipped with adaptive $K_{\mathrm{tr}}$ adjusted by accumulator value $\gamma$, can suppress overshoots during the irradiance recovery process.

\begin{figure}[htb]
	\centerline{\includegraphics[width=0.4\textwidth
	]{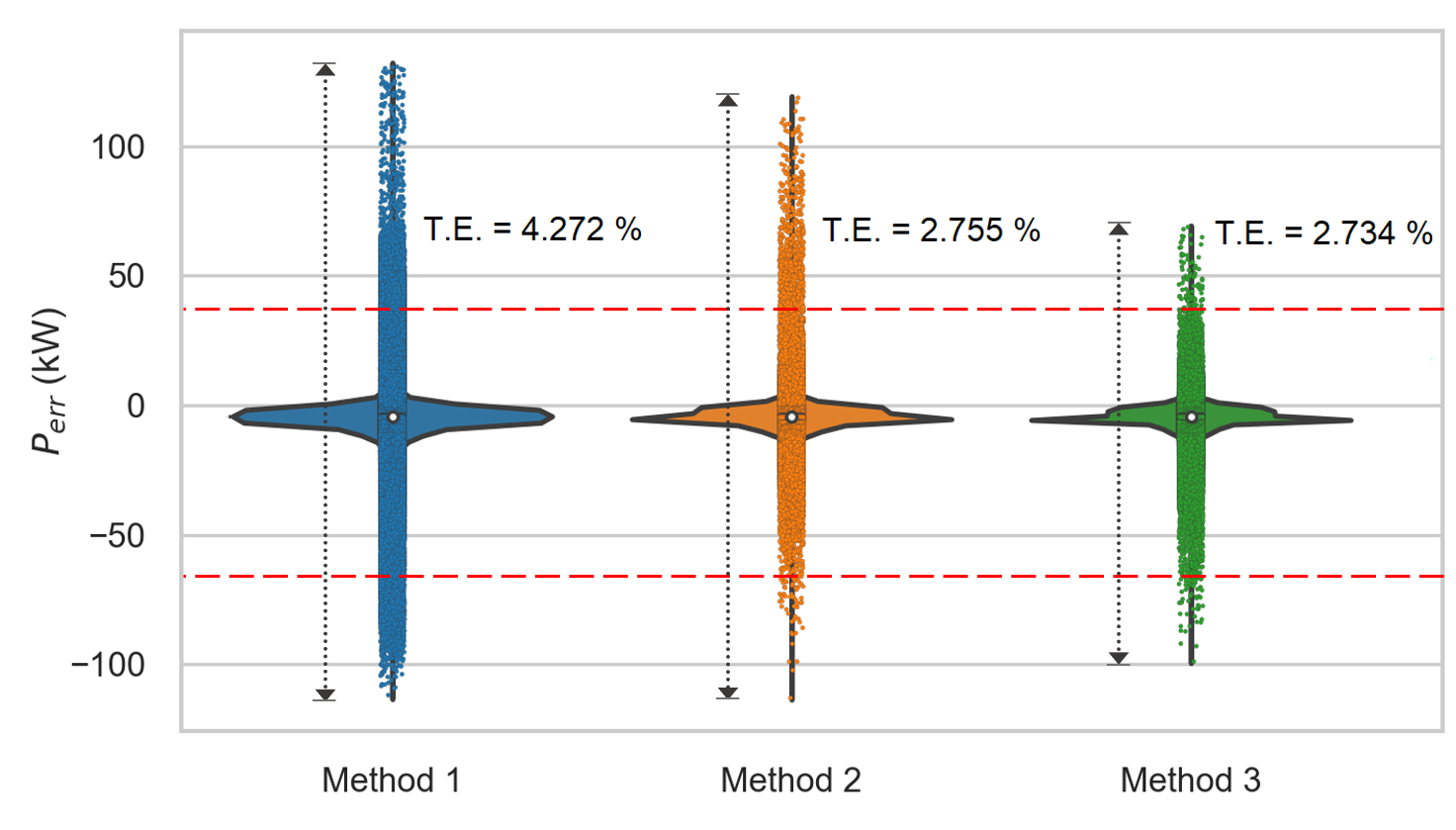}}
	\caption{Tracking error comparison.}
	\label{violin}
\end{figure}

Tracking errors with a 0.1s resolution are calculated for three days of operation. When $P_\mathrm{{ref}}\leq P_\mathrm{{mpp}}$, the error is calculated as $P_\mathrm{{err}}=|P_\mathrm{{pv}}-P_\mathrm{{ref}}|$; when $P_\mathrm{{ref}}>P_\mathrm{{mpp}}$, the error is calculated as $P_\mathrm{{err}}=|P_\mathrm{{pv}}-P_\mathrm{{mpp}}|$. The cumulative percentage of tracking error, $E_\mathrm{sum}$ is calculated by  
\begin{equation}
    E_\mathrm{sum} = \frac{\int|P_{err}|}{\int|P_{pv}|}
\end{equation}
Figure \ref{violin} displays violin plots overlaid by scatter plots of $P_\mathrm{{err}}$ for the three methods. The dashed red lines indicate the region of concentrated data points from scatter plot of method 3. It can be seen that the error range of method 3 is much smaller compared to methods 1 and 2. Furthermore, in this analysis, the adaptive gains in methods 2 and 3 are kept the same for a fair comparison. Yet, the real-time adjustment of the adaptive gain introduced in method 3 permits its system to be designed with a higher adaptive gain base value, which can be leveraged to reduce the tracking error even further.

\section*{Acknowledgment}
The authors thank PJ Rhem with ElectriCities, Paul Darden, Steven Hamlett and Daniel Gillen with Wilson Energy for their inputs, suggestions and technical support.  

\section{Conclusion}
In this paper, we presented an algorithm for adjusting the dc-link voltage of a PV system to achieve a better power reference tracking capability. The algorithm added two new mechanisms for adjusting the dc-link voltage. First, by comparing the PV power output with its average output, it can identify when the MPP is below the power reference. Then, the gain to adjust voltage step size can be reduced in real-time so that it is only large when needed. Second, an overshoot detection mechanism is used to trigger an accumulator for suppressing overshoots. Simulation results demonstrate a dc-link with reduced oscillations in lower-irradiance operation conditions, and an improved overshoot response when compared to existing P\&O-based methods in the literature. 

\bibliographystyle{IEEEtran}
\bibliography{mybibtex}

\end{document}